\documentclass[useAMS,usenatbib]{mn2e}
\usepackage{rotating}
\usepackage{lscape}

\def\vsini{$v$sin$i$~}
\def\teff{$T_{\rm{eff}}$~}
\def\logg{log~$g$~}

\title[Age and Abundances of PZ Tel]{Benchmark cool companions:  Ages and abundances for the PZ Tel system}
\author[J.S. Jenkins et al.]{J.S. Jenkins$^{1}$\thanks{E-mail:
jjenkins@das.uchile.cl}, Y.V. Pavlenko$^{2,3}$, O. Ivanyuk$^{2}$, J. Gallardo$^{1}$, M.I. Jones$^{1,4}$, \newauthor
A.C. Day-Jones$^{1}$, H.R.A. Jones$^{3}$, M.T. Ruiz$^{1}$, D.J. Pinfield$^{3}$ and L. Yakovina$^{2}$\\
$^{1}$Departamento de Astronomia, Universidad de Chile, Camino el Observatorio 1515, Las Condes, Santiago, Chile, Casilla 36-D\\
$^{2}$Main Astronomical Observatory, Academy of Sciences of Ukraine, Golosiiv Woods, Kyiv-127, 03680 Ukraine\\
$^{3}$Center for Astrophysics Research, University of Hertfordshire, College Lane, Hatfield, Herts, UK, AL10 9AB\\
$^{4}$European Southern Observatory, Casilla 19001, Santiago, Chile}
\begin{document}

\date{Submitted January 2011}

\pagerange{\pageref{firstpage}--\pageref{lastpage}} \pubyear{2002}

\maketitle

\label{firstpage}

\begin{abstract}

We present new ages and abundance measurements for the pre-main sequence star PZ Tel.  PZ Tel was recently 
found to host a young and low-mass companion.  Such companions, whether they are brown dwarf or planetary, can attain benchmark 
status by detailed study of the properties of the primary, and then evolutionary and bulk characteristics can be inferred for the 
companion.  Using FEROS spectra we have measured atomic abundances (e.g. Fe and Li) and chromospheric activity for PZ Tel and used these 
to obtain metallicity and age estimates for the companion.  We have also determined the age independently using the latest evolutionary 
models.  We find PZ Tel to be a rapidly rotating (\vsini=73$\pm$5kms$^{-1}$), approximately solar metallicity star (log$N$(Fe)=-4.37~dex or 
[Fe/H]=0.05~dex).  We measure a NLTE lithium abundance of log$N$(Li)=3.1$\pm$0.1~dex, 
which from depletion models gives rise to an age of 7$^{+4}_{-2}$~Myrs for the system.  Our measured 
chromospheric activity (log$R'_{\rm{HK}}$ of -4.12) returns an age of 26$\pm$2~Myrs, as does fitting pre-main sequence evolutionary 
tracks ($\tau_{evol}$=22$\pm$3~Myrs), both of these are in disagreement with the lithium age.  We speculate on reasons for this difference 
and introduce new models for lithium depletion that incorporate both rotation and magnetic field affects.  
We also synthesize solar, metal-poor and metal-rich substellar evolutionary models to better determine the bulk properties of PZ Tel~B, showing that PZ Tel~B is 
probably more massive than previous estimates, meaning the companion is not a giant exoplanet, even though a planetary-like formation origin can go 
some way to describing the distribution of benchmark binaries currently known.  We show how PZ Tel~B compares 
to other currently known age and metallicity benchmark systems and try to empirically test the effects of dust opacity as a function of metallicity 
on the near infrared colours of brown dwarfs.  Current models suggest that in the near infrared observations are more sensitive to low-mass companions orbiting 
more metal-rich stars.  We also look for trends between infrared photometry and metallicity amongst a growing population of substellar 
benchmark objects, and identify the need for more data in mass-age-metallicity parameter space.

\end{abstract}

\begin{keywords}

stars: abundances, stars: activity, stars: chromospheres, stars: low-mass, brown dwarfs, stars: pre-main sequence, stars: planetary systems

\end{keywords}

\section{Introduction}

PZ Telescopii (HD174429), more commonly known as PZ Tel, is classed as a G9IV star and is found to be a variable star of the BY Draconis type.  The star has 
been studied by many authors since it is a nearby and bright pre-main sequence star.  In particular, PZ Tel has been a major focus for work relating to stellar activity 
and differential rotation (e.g. \citealp{bopp83}; \citealp{innis88}; \citealp{barnes00}), it is thought to be part of the $\beta$~Pictoris Moving Group, hereafter BPMG, 
(\citealp{zuckerman01}), and it has been shown to host a debris disk of remnant formation material of around 0.3~Lunar-masses that spans a radius of 165~AU, 
with the inner edge located only 35~AU from the central star (\citealp{rebull08}).

More recently, PZ Tel has been found to be part of a binary system with a low-mass brown dwarf companion \citep{biller10}.  Biller et al. used the Gemini-NICI adaptive optics 
system to discover the faint and co-moving companion PZ Tel~B and from their analysis of its colours and luminosity, they found it to have a mass of 36$\pm$6M$_{\rm{J}}$, one 
of the lowest mass binary companions yet detected.  Soon after this discovery, \citet{mugrauer10} confirmed the existence of PZ Tel~B using observations made with the VLT-NACO 
system.  They find a mass for the companion of 28$^{+12}_{-4}$M$_{\rm{J}}$, assuming an age of 12$^{+8}_{-4}$Myrs and using the evolutionary models of \citet{chabrier00} and 
\citet{baraffe02,baraffe03}, in good agreement with the mass found by Biller et al.  The physical separation of the pair is found to be $\sim$16~AU at present, locating the 
companion within the inner edge of the debris disk, which was probably cleared out by the formation and evolution of PZ Tel~B.

Binary systems that contain a brown dwarf companion and a host star that can give rise to robust evolutionary or physical parameters are very useful tools as calibrators for 
modeling the physics of substellar atmospheres.  Main sequence stars as hosts can provide very precise metallicities for brown dwarf models to benchmark their efforts.  \citet{liu07} 
imaged a faint T-dwarf companion to the exoplanet host star HD3651A, allowing a precise metallicity to be assumed for the T-dwarf.  However, robust age estimates for field main sequence 
stars are generally difficult to acquire since the evolutionary tracks tend to converge on the main sequence.  Additionally, main sequence field stars 
tend to be very old, which makes it difficult to image low-mass companions that are even known to exist (see \citealp{jenkins10}).  However, a number of surveys are underway to 
detect more brown dwarf binary companions (see \citealp{pinfield05}), which have led to recent discoveries (\citealp{day-jones08}; \citealp{zhang10}) 
including the discovery of the first T-dwarf companion orbiting any white dwarf primary star (\citealp{day-jones11}).

\subsection{Planetary Origin?}

Another possibility is that PZ Tel~B is actually a directly imaged exoplanet.  
30\% of disk material can be locked up into forming planets (\citealp{mordasini08}).  The maximum stable disk mass is given by Md = 0.1 $\times$ M$_{\odot}$.
Therefore, the maximum mass of material that can be processed into forming rocky cores is given by Mz = Percentage of material $\times$ Md $\times$ $Z$ $\times$ (3.3x10$^5$).  
For instance, a solar mass star can contain Mz = 0.3 $\times$ 0.1 $\times$ 1.0 $\times$ 0.02 $\times$ 3.3x10$^5$ $\simeq$ 200M$_{\oplus}$ of rocky material that 
can form planetesimals.  However, for PZ Tel~A, the disk mass Md would have been 0.10 $\times$ 1.16 $\sim$ 0.12M$_{\odot}$, and our metallicity measurement 
gives rise to a solar $Z$ fraction of $\sim$2\%.  Inputing these numbers means that the remnant disk around PZ Tel~A would have contained 0.3 $\times$ 1.2 $\times$ 0.10 $\times$ 0.02 
$\times$ 3.3x10$^5$ $\simeq$ 240M$_{\oplus}$ of material that could be processed into forming cores.  Given that the current PZ Tel debris disk only contains 0.3~Lunar-masses 
of material, the majority of the mass of metals could have gone into forming a huge core of a few hundred Earth-masses, large enough to form a planet like the 
observed companion PZ Tel~B; or in this case PZ Tel~b.

This scenario is in agreement with the latest population synthesis models (\citealp{mordasini09a}) where very 
massive planets (20-40M$_{\rm{J}}$ or so) can form through core accretion, but only in a handful of cases.  \citet{sahlmann10} also suggest the dividing line between massive planetary 
companions and brown dwarf companions resides in the range between 25$-$45~M$_{\rm{J}}$'s.  The current mass estimates for the PZ Tel companion would place it well 
within this region.  It stands to reason therefore that the PZ Tel system could be the first system where we have directly imaged one of these new extreme-Jovian planets.

\section[obs]{Observations and Methodology}

\subsection{FEROS Spectra}

Two PZ Tel spectra were observed using the Fibre-fed Extended Range Optical Spectrograph (FEROS; \citealp{kaufer99}) on the MPG/ESO - 2.2m telescope in 2007 and 2010.  S/N ratios of 
over 100 in the continuum at 7500\AA\ and $\sim$50-60 at the Ca\sc~ii\rm~HK lines (3955\AA) were obtained and FEROS maintains a resolving power of 
$R\sim$48$'$000.  The reduction procedure for all spectra is described in more detail in \citet{jenkins08,jenkins11}.  All data were debiased, 
flatfielded, had the scattered-light removed, optimally extracted, and had the blaze function removed using the FEROS pipeline and a number of Starlink procedures.  

\subsection{Metallicity Determination}

\begin{figure}
\vspace{5.5cm}
\hspace{-4.0cm}
\includegraphics{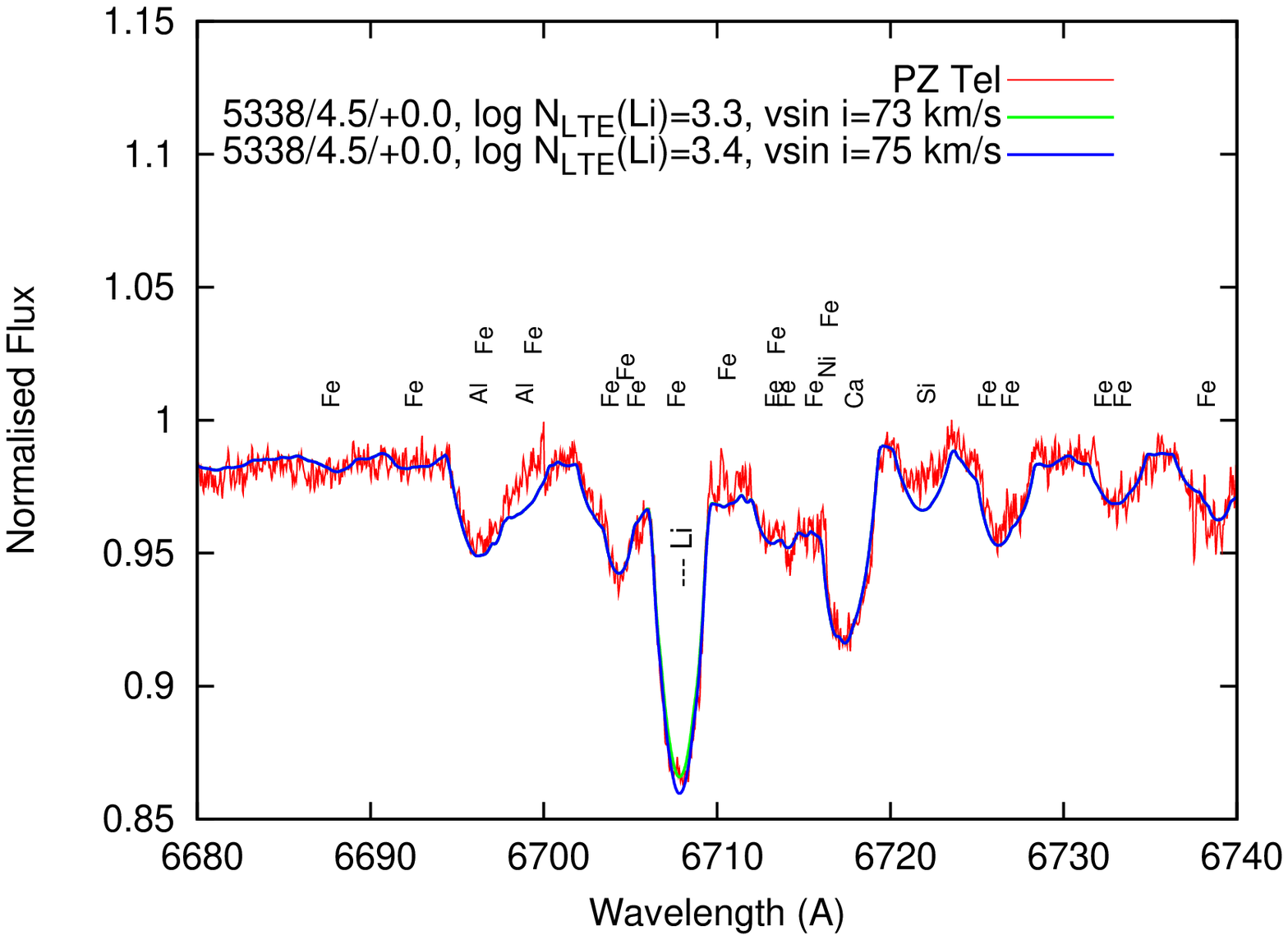}
\includegraphics{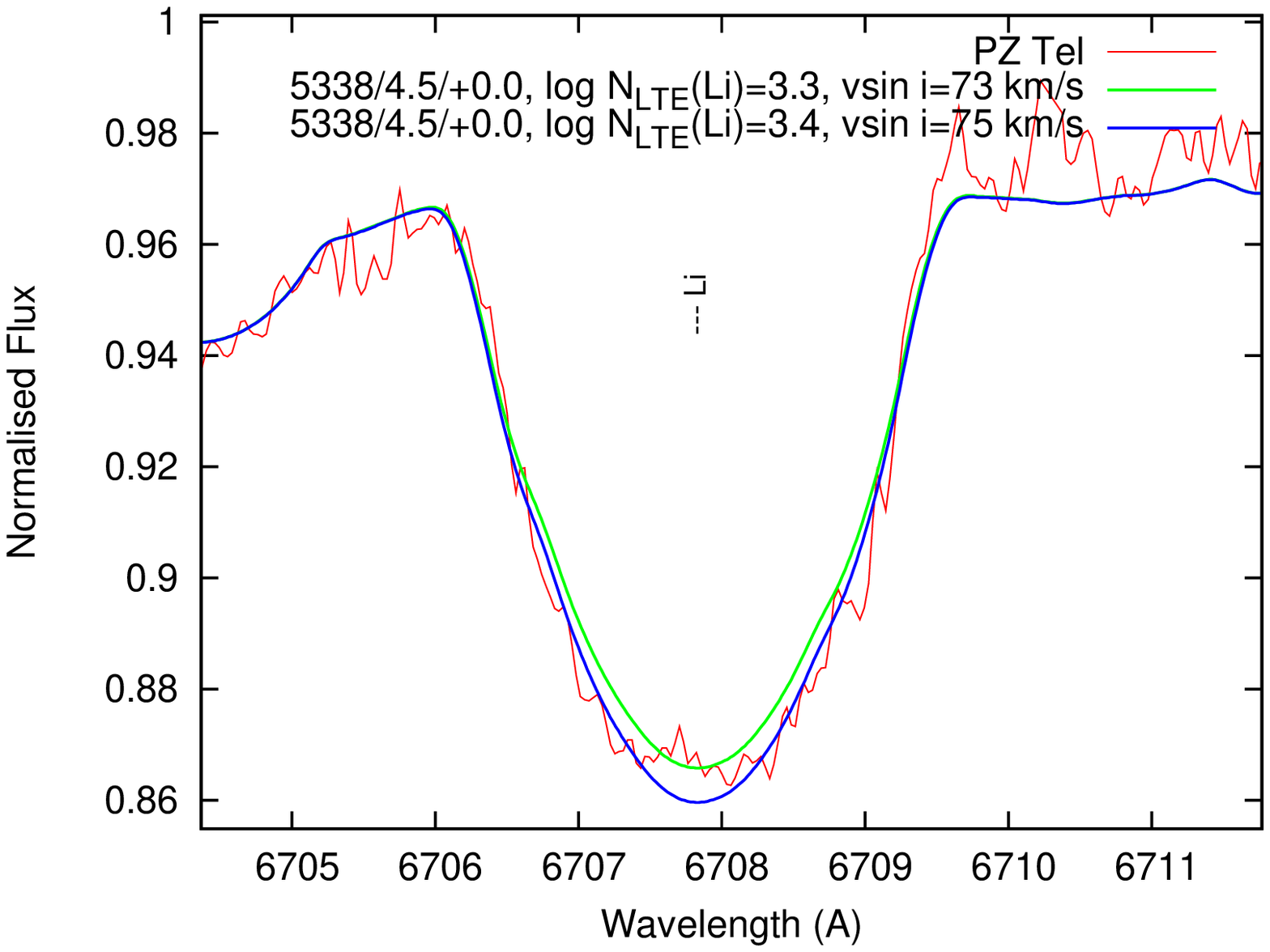}
\vspace{6.0cm}
 \caption{A subsection of the FEROS observed spectrum of PZ Tel (red) is shown in a part of our region of interest.  The top panel shows a wide region and highlights a number of 
iron lines, the lithium lines and other species like aluminium, calcium, and silicon.  Overplotted in green is our best fit synthetic spectrum with the input parameters shown in the key 
and the blue model is a comparison model of higher abundance.  The lower panel is a zoomed in plot of the lithium lines and shows the same two LTE model spectra as in the top panel.}
\label{pztelspec}
\end{figure}

Our metallicities were determined using a similar method to that explained in detail in \citet{pavlenko11} except with a few modifications to take into consideration the 
high rotational velocity (\vsini) of PZ Tel, a method that we will outline in a companion paper (\citealp{ivanyuk11}).  We first started with a high resolution and high S/N 
\citet{kurucz84} solar spectrum as our template star and then broadened this spectrum to 70~kms$^{-1}$ 
to closely match the spectrum of PZ Tel.  We used WITA6 (\citealp{pavlenko97}) to synthesize all spectra by computing plane-parallel, and self consistent, model atmospheres 
using SAM12 (\citealp{pavlenko03}), along with a list of atomic and molecular lines drawn from VALD2 (\citealp{kupka99}) with some updates from \citet{yakovina11}.  We then 
determined the best lines/regions in the 
broadened solar spectrum that would give rise to the well known solar metallicity value using our spectral synthesis fitting procedure.  Once these lines were selected, we 
then select the same regions in the PZ Tel spectrum and perform the same analysis as we did on the solar spectrum to get the iron abundance of PZ Tel.

Fig.~\ref{pztelspec} shows the results from our fitting procedure.  We show best fits in one of our regions of interest, both over a wide spectral range (top panel) and a zoomed in 
region around the lithium lines (lower panel).  In these plots we see two LTE model spectra fits (green and blue), with different abundances, against the 
observed spectrum of PZ Tel (red).  Also shown is the model atmosphere we have used to synthesize these spectra.  In this region we also take into account lines of CN and aluminium.  
It can be seen that 
the models fit the data well, considering the difficulties that one encounters by processes such as continuum fitting rotationally broadened spectra like these, and so we believe 
that our measured metallicity is robust.  

Interestingly, we find the Fe abundance of PZ Tel to be log$N$(Fe)=-4.37$\pm$0.06~dex ([Fe/H]=+0.05$\pm$0.20~dex) meaning it is a solar/slightly metal-rich, young star, a result that is 
not consistent with most previous metallicity measurements (e.g. \citealp{rocha-pinto98}).  
However, we have tested our method on a model young Sun, fit to more than one Fe line, and used a methodology we have tested on Sun-like stars in the past that was 
shown to be insensitive to small variations in the input parameters like \teff and \logg ~(\citealp{pavlenko11}).  Therefore, our value should be more robust than those measured 
previously. 

 Table~\ref{met_test} shows the sensitivity of our results over a range of \teff values running from $\pm$300K around the measured \teff of PZ Tel~A in the literature (e.g. \citealp{randich93}).  
The abundances range from -4.60 up to -4.25 in log~$N$(Fe), which relates to a range in [Fe/H] of -0.18 up to +0.17~dex.  This analysis 
shows that over a span of 600K, we find changes in the metallicity of $\pm$0.18~dex, standard deviation of $\pm$0.13, and we also see from the table that the uncertainties 
on the individual fits are similar across all \teff's, with a little lower uncertainties for the higher \teff models.  
We studied the relationship between the 
excitation energy of the individual lines against their measured abundance, and a trend may be present in the data at a \teff of 5238K, however we note that most of our lines have 
similar excitation energies and only one line had a significantly lower one, which determines the slope of the fit.  Yet the slope did appear to correlate with changing \teff as expected.  
Therefore, we believe the \teff of PZ Tel~A to be slightly hotter than the previous estimates of 5238K, and so we adopt 5338$\pm$200K, giving rise to our measured abundance of 
+0.05$\pm$0.2~dex, in good agreement with a solar metallicity for the star.  A hotter \teff can also found from the evolutionary tracks if the conversion table from \citet{kenyon95} is 
employed with the \citet{siess00} evolutionary models (see next section). 

\begin{table}
\small
\center
\caption{Calculated metallicity values for different temperatures around the literature value of 5238 K found by \citet{randich93}.}
\label{met_test}
\begin{tabular}{ccc}
 & & \\
\hline 
\multicolumn{1}{c}{\teff [K]}  & \multicolumn{1}{c}{log$N$(Fe) [dex]} & \multicolumn{1}{c}{[Fe/H] [dex]}   \\
\hline 
 4938 &  -4.60$\pm$0.07 &  -0.18 \\
 5038 &  -4.53$\pm$0.07 &  -0.11  \\
 5138 &  -4.49$\pm$0.07 &  -0.07 \\
 5238 &  -4.43$\pm$0.06 &  -0.01  \\
 5338 &  -4.37$\pm$0.06 &  +0.05 \\
 5438 &  -4.30$\pm$0.05 &  +0.12 \\
 5538 &  -4.25$\pm$0.05 &  +0.17 \\
\hline 
\end{tabular}
\medskip
\end{table}

Finally, given that our Fe lines are located in the optical part of the spectrum, and PZ Tel~B is 5.04 magnitudes fainter than PZ Tel~A in the K$_{s}$-band, where low-mass 
brown dwarfs emit much more flux than in the optical, we can be fairly sure that changes in the metallicity measurements for PZ Tel~A between different authors is not due to varying 
levels of contamination from the secondary component.

\begin{table*}
\small
\center
\caption{PZ Tel calculated values}
\label{pztel_table}
\begin{tabular}{ccccccc}
 &&&&&& \\
\hline 
\multicolumn{1}{c}{Star}  & \multicolumn{1}{c}{V}  & \multicolumn{1}{c}{$B-V$}  & \multicolumn{1}{c}{\teff}  & \multicolumn{1}{c}{\logg} & \multicolumn{1}{c}{M$_{\star}$} & 
\multicolumn{1}{c}{R$_{\star}$}  \\
\multicolumn{1}{c}{}  & \multicolumn{1}{c}{(mags)}  & \multicolumn{1}{c}{}  & \multicolumn{1}{c}{(K)}  & \multicolumn{1}{c}{(dex)} & \multicolumn{1}{c}{M$_{\odot}$} & 
\multicolumn{1}{c}{R$_{\odot}$} \\ \hline

HIP92680~A          &  8.43 &  0.784 & 5338$\pm$200 & 4.41$\pm$0.10 & 1.13$\pm$0.03 & 1.23$\pm$0.04 \\
\\
\hline
\multicolumn{1}{c}{[Fe/H]}  & 
\multicolumn{1}{c}{log$R'_{\rm{HK}}$}  & \multicolumn{1}{c}{\vsini} & \multicolumn{1}{c}{log~$N$(Li)}  & \multicolumn{1}{c}{$\tau$$_{evol}$}  & \multicolumn{1}{c}{$\tau$$_{gyro}$}  & 
\multicolumn{1}{c}{$\tau$$_{Li}$} \\
\multicolumn{1}{c}{(dex)}  & 
\multicolumn{1}{c}{(dex)}  & \multicolumn{1}{c}{(kms$^{-1}$)}  & \multicolumn{1}{c}{dex}  & \multicolumn{1}{c}{(Myrs)}  & \multicolumn{1}{c}{(Myrs)}  & \multicolumn{1}{c}{(Myrs)} \\ \hline

+0.05$\pm$0.20 & -4.12$\pm$0.06 & 72$\pm$5 & 3.1$\pm$0.1 & 22.34$\pm$3.13 & 26.23$\pm$1.62 & 7$^{+4}_{-2}$  \\

\hline 
\end{tabular}
\medskip
\end{table*}

\section{Age Estimations}

\subsection{Evolutionary Age}

We first derive the age of PZ Tel~A by plotting the star on an HR-diagram using the latest Hipparcos data available from \citet{perryman97} and \citet{vanleeuwen07}.  
From Hipparcos we find the star to have a $B-V$ colour index of 0.784 and, with a parallax of 19.42$\pm$0.98mas, we obtain a distance of 51.49$\pm$2.60pc.

We find that PZ Tel~A is most likely a pre-main sequence star, yet to reach the zero age main sequence. 
We interpolate its position onto the CESAM (\citealp{marques08}) and \citet{siess00} isochrones and isomass tracks in a similar manner to that in \citet{jenkins09a}.  
The CESAM models give rise to a mass of 1.14$\pm$0.03M$_{\odot}$, a surface gravity (\logg) of 4.41$\pm$0.10~dex, radius of 1.26$\pm$0.04R$_{\odot}$ and an age 
of 22.17$\pm$2.05Myrs.  These are in good agreement with the values from the Siess et al. evolutionary models which also take into consideration the metallicity of the star.  The mass, 
radius and age from these models are 1.11$\pm$0.03M$_{\odot}$, 1.20$\pm$0.03R$_{\odot}$ and 22.51$\pm$2.36Myrs, respectively.  The final values are listed in Table~\ref{pztel_table}.

\subsection{Gyrochronological Age}

We extract the FEROS chromospheric log$R'_{\rm{HK}}$ activity at two epochs for PZ Tel, once in 2007 and again in 2010.  This gives a fairly robust activity measurement 
however the activity cycle is expected to be fairly large therefore we may still be miss representing the true mean activity level.  We measured these activities following the 
same procedure explained in \citet{jenkins06c,jenkins08}.  Briefly, the Ca\sc~II~\rm~HK lines, located at 3968.470\AA\ and 3933.668\AA\ respectively, were filtered through two triangular 
bandpasses centered on their line cores and with FWHM's of 1.09\AA.  We then compare the ratio of these filtered fluxes to another two filtered square bandpass regions in the 
surrounding continuum, centered at 3891\AA\ and 4001\AA, labelled the V and R bands respectively.  This ratio is highlighted in Eqn.~\ref{eq:activity} where the N$_{i}$ is the 
integrated fluxes in each filtered bandpass region.

\begin{equation}
\label{eq:activity}
N_{FEROS} = \frac{N_{H} + N_{K}}{N_{V} + N_{R}}
\end{equation}

This ratio is then normalised to the bolometric luminosity of the host star to extract the chromospheric part of the spectral light and using the relations from \citet{noyes84a} 
we arrive at the final log$R'_{\rm{HK}}$ activity index.  For PZ Tel we find log$R'_{\rm{HK}}$ activities of -4.16 and -4.07~dex for the 2007 and 2010 data respectively.  The mean of 
these measurements (-4.12$\pm$0.06~dex) is used to derive its age.  The difference between the two values highlights the 
young and active nature of the star, given the uncertainties are only at the level of $\pm$0.02~dex for these observations. 

In order to obtain an age estimate from the activity of PZ Tel we use the latest age-activity relations from \citet{mamajek08}.  The Mamajek \& Hillenbrand relation gives rise to 
a mean age of 26.23$\pm$1.62~Myrs for PZ Tel, with a range between 17$-$40~Myrs for the two individual measurements.  The uncertainties on the age estimation were taken 
using the published scatter around the age-activity fit, along with the uncertainties on the mean age derived from the two observations.  This age estimation is in very good agreement 
with the age we have derived from the isochrone fitting procedure above.

\subsection{Lithium Age}

\begin{figure}
\vspace{5.5cm}
\hspace{-4.0cm}
\includegraphics{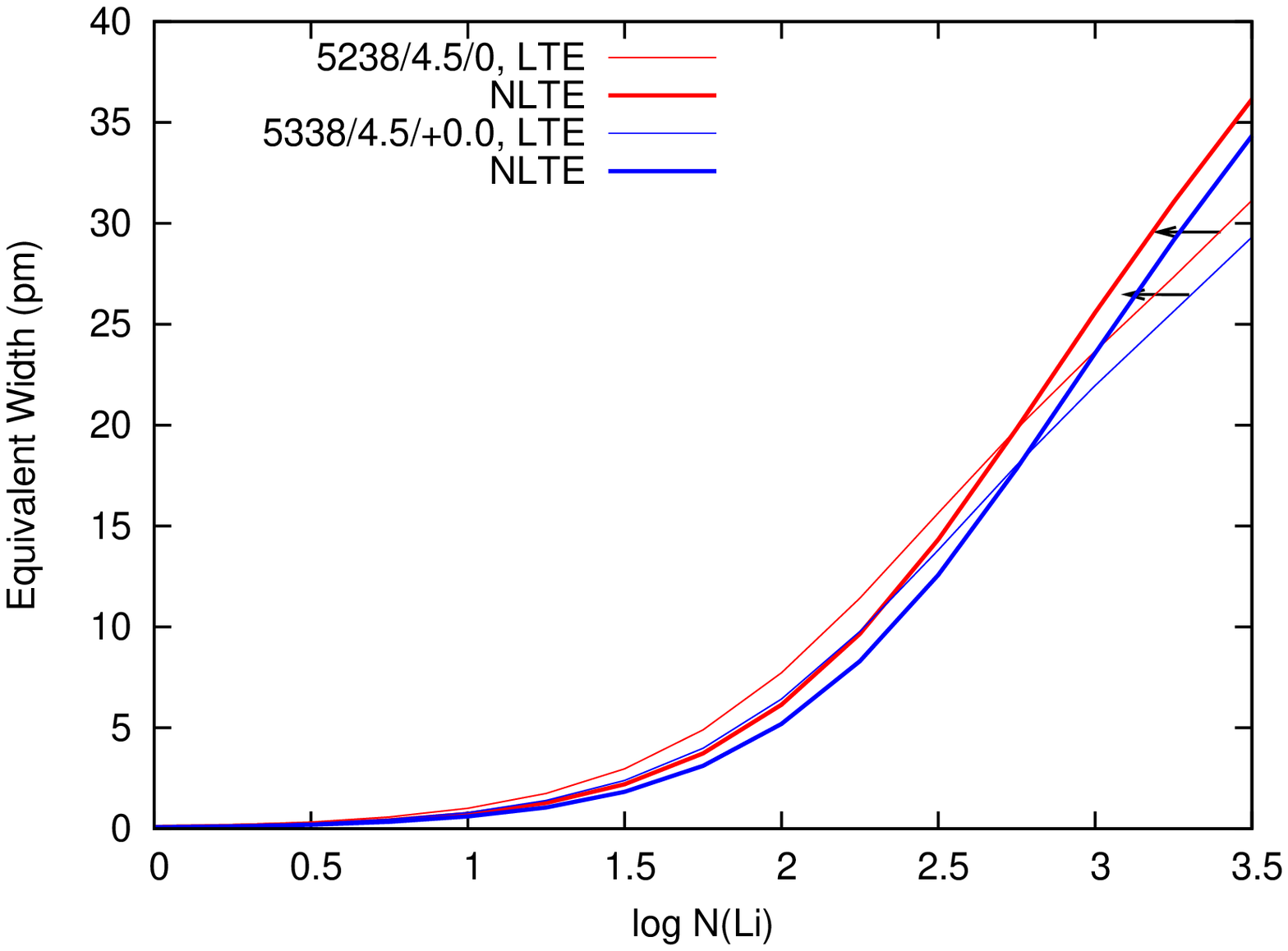}
\includegraphics{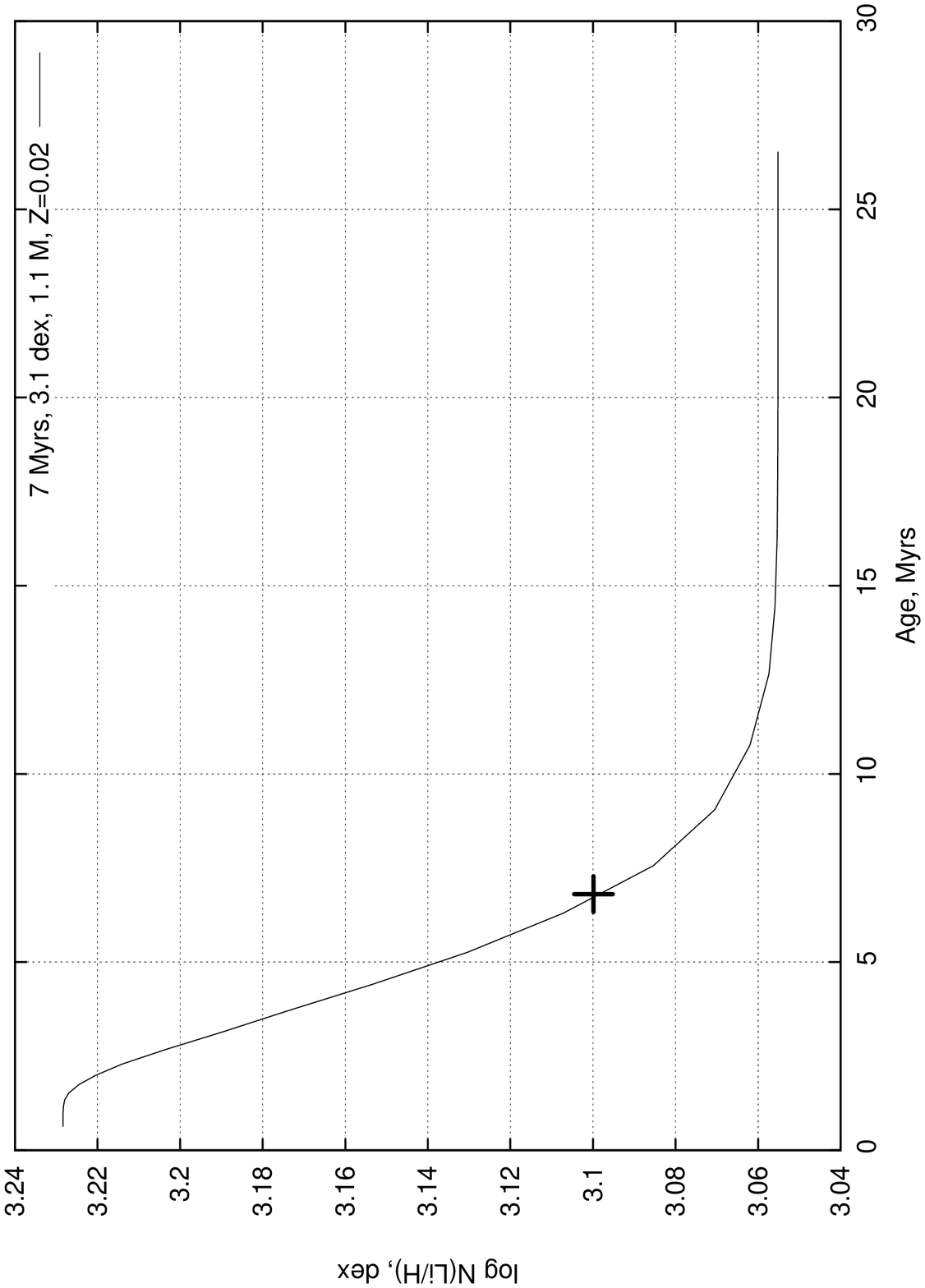}
\vspace{6.0cm}
 \caption{Top panel shows the differences in measured equivalent width of lithium by adopting LTE (solid curves) or NLTE (dashed curves).  The horizontal arrow shows the measured values for PZ 
Tel~A for two plausible \teff values.  The lower panel shows lithium depletion as a function of age taken from the \citet{siess00} pre-main sequence models for a 1.15M$_{\odot}$ star.  The 
cross represents our measured value for PZ Tel~A and the key in top right highlights the measured age, lithium abundance, and metal fraction employed.}
\label{lithium}
\end{figure}

We also measure an age for PZ Tel through model fitting the Li lines at 6708\AA\ (as shown in Fig.~\ref{pztelspec}).  Strong lithium absorption is another indicator of youth in 
stars since lithium is destroyed at temperatures above around 2.5x10$^6$K, and therefore this element is depleted through time in the interiors of stars.  The depletion of lithium is dependent 
on a number of factors that influence the convective envelopes of stars and the mixing processes and convection therein (see \citealp{pinsonneault97}).  Yet lithium can still be used as 
a good indicator of the age of young stars, even though some studies have shown that evolutionary models may under-predict the lithium depletion, giving rise to systematically 
older ages (\citealp{zuckerman01}; \citealp{white05}).  

In the lower panel of Fig.~\ref{pztelspec} we see our best fit synthetic spectra to the lithium region.  The models we show fit the 
observed spectrum well and we find the best fit to be for a \vsini of 73~kms$^{-1}$ and a LTE lithium abundance of log~$N_{\rm{LTE}}$(Li) 3.3.  We then follow the NLTE computational 
procedure explained in \citet{pavlenko96} to arrive at our final value for the lithium abundance of 
PZ Tel of log~$N_{\rm{NLTE}}$(Li) 3.1$\pm$0.1~dex, in good agreement with that found by previous authors (e.g. \citealp{randich93}; \citealp{soderblom98}).  

We highlight that our value is drawn from a NLTE analysis, since the metallicity we measure for PZ Tel~A was found under the assumption of LTE.  The top panel in Fig.~\ref{lithium} shows the 
difference in equivalent width of the lithium line as a function of measured abundance.  The LTE curves have a shallower gradient at a higher log~$N$(Li) than the NLTE curves and both cross 
at a log~$N$(Li) of around 2.6-2.8.  We show two different temperature tracks and we see that in general, models of higher \teff provide higher lithium abundances.  Our LTE spectral 
synthesis provides the best fit Li abundance of 3.3 for the lithium abundance, shown by the horizontal arrow in the figure, but this value drops to our measured value of 3.1 when 
examined in NLTE.

Fig.~\ref{lithium} (lower panel) shows the lithium depletion evolution models from \citet{siess00} for the measured mass of PZ Tel~A (1.15M$_{\odot}$).  Lithium models generally are for non-rotating 
stars, whereas young and fast rotators like PZ Tel induce processes difficult to model accurately.  Our measured lithium abundance for PZ Tel is highlighted on this plot by the cross.  We also 
show the age and abundance value in the key at top right of the plot.  We measure a lithium depletion age for PZ Tel of only 7$^{+4}_{-2}$~Myrs, younger than both the chromospheric age and the 
evolutionary model age.  Also, the chromospheric age and evolutionary model fitting ages are far older than the depletion timescale for a star at this \teff by a factor of two or so.  
Such discrepancies require explanation, particularly since the evolution of \teff as a function of age from the same models yield an age of 23~Myrs.

\subsubsection{Rotation, Magnetic Fields and Accretion}

\begin{figure}
\vspace{5.0cm}
\hspace{-4.0cm}
\includegraphics{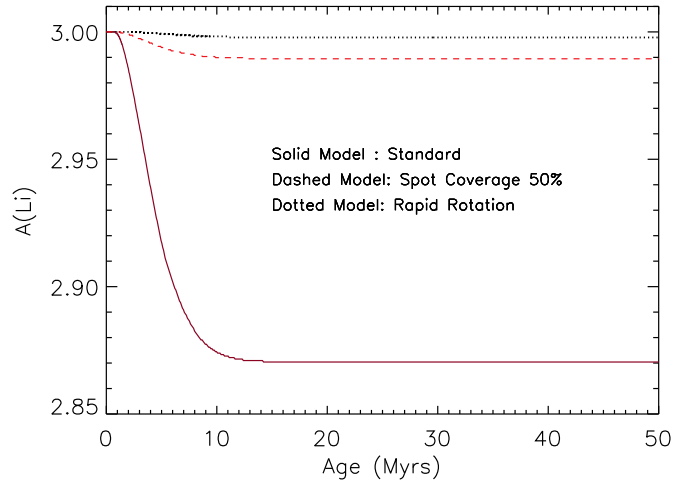}
\vspace{1.7cm}
\caption{Three 1.15M$_{\odot}$ models of lithium depletion for a standard, non-rotating, and non-magnetic star (solid), a magnetic star (dashed), and a fast rotating star (dotted).}
\label{lith_rot_mag}
\end{figure}

As mentioned, the depletion models are non-rotating models, whereas PZ Tel~A is a fast rotating star.  We could envisage that significant rotation could drive powerful magnetic fields 
(\citealp{kraft67}), which could inhibit the convective motions within the young stellar atmosphere.  Inhibiting this motion could lead to reduced lithium depletion in the interior 
of the star (\citealp{martin94}).  In addition, these magnetic fields could also reduce the level of differential rotation in these stars, yielding less shearing and again further 
reduced lithium depletion.  \citet{barnes00} has studied the level of differential rotation in PZ Tel~A and found the surface shear to be similar to the solar shear, possibly 
supporting this scenario, but this is also similar to other young stars like Speedy Mic (\citealp{barnes01}).

We have investigated this issue using new models currently under construction based around the explanations presented in \citet{chabrier07} who introduced activity effects 
in low-mass stellar models by considering two parameters: the spot blocking factor ($\beta$) and the modification of the mixing length parameter ($\alpha$). This study shows that the effects of 
the $\alpha$ and $\beta$ parameters are degenerate, i.e. the properties of any given system can be reproduced by modifying any of the two or both.

The authors also showed that the introduction of rotation and/or magnetic field effects on the models of \citet{baraffe98} could explain the discrepancies between the observed and theoretically 
predicted mass-radius relationship of eclipsing binaries. Specifically, they presented two scenarios considering: (1) that the effect of magnetic fields and rotation alter the efficiency of 
convective energy transport, which can be modelled by setting the mixing length parameter (Mixing Length Theory) to lower values than those used for solar models; and (2) that the stellar 
magnetic activity present on these objects can be associated with the appearance of dark spots covering the radiative surface, modeled by the $\beta$ factor indicating the percentage of the 
stellar surface covered by spots.

The results of \citet{chabrier07} show that both these scenarios predict larger radii than standard stellar models, but while the effects of spots are significant over the entire low-mass domain, 
the effect on convection is relatively small for fully convective stars. Moreover, modified evolutionary models present cooler central temperatures, affecting burning rates of light elements.
Specifically, lithium is depleted more slowly in models where the convection is inhibited or where spots are present.  If the lithium is burning at different rates, the age derivation for 
stellar or substellar objects needs to be revised due to the effects of activity or rotation in this calculation.  Work is under way to explore these effects (\citealp{gallardo11}). 

Fig.~\ref{lith_rot_mag} shows preliminary
results for the Li evolution of a 1.15M$_{\odot}$ object for a standard model (solid line, spot-free and $\alpha$=1.9), and two "active" models: short-dash line ($\alpha$=1.9
and $\beta$=0.5 i.e., 50 \% surface coverage by spots) and dotted line (fast rotation, $\alpha$=0.5 and spot-free).  As we can see from the plot, the "active" models significantly decrease the rate 
of lithium depletion, through inhibiting the convective motions in the stellar interior.  Thus, rotation and/or magnetic fields (through spot coverage) could be present on our target and could 
give rise to the high Li abundance we measure for PZ Tel~A, meaning a younger age from a non-rotating model could be expected.

An additional possibility is that the lithium depletion is not being affected at all by magnetic fields and high rotation, but that the level of lithium is 
being replenished somehow in the atmosphere of PZ Tel~A.  Given that we find PZ Tel~A to have around solar metallicity we expect there was a lot of dust in the proto-planetary disk and 
hence planetesimal formation was a strong possibility.  Accretion of planetesimals could significantly replenish the level of lithium in the atmosphere of PZ Tel~A, giving rise to the 
measured value of 3.1 that we currently find for this star.  

\citet{israelian01,israelian03} suggest that the planet host star HD82943 has engulfed a planet of around 
2M$_{\rm{J}}$, leading to an enhancement of the $^6$Li isotope up to the measured value of 4.5x10$^{44}$ nuclei, therefore a similar process could explain the overabundance of lithium in PZ Tel~A's 
atmosphere.  For such young and rapidly rotating stars it is difficult to conclude this is the case since the uncertainty on the lithium abundance is large.  

In addition, \citet{baraffe10} have shown that episodic accretion onto the star can significantly affect the lithium abundance, adding another source of uncertainty.  Given these results and 
conclusions, we do not use the age derived from the lithium abundance in our final mean age for the PZ Tel system.

Taken all together, we get an average age of 24$\pm$3~Myrs for PZ Tel, which is slightly higher than the previous estimates given that PZ Tel is thought to be a member of the 
BPMG (\citealp{zuckerman01}), but only at the level of 1$\sigma$.  However, this could indicate that PZ Tel is not a bonafide member of the BPMG, or that the true age of the BPMG is 
actually significantly older than the 12$^{+8}_{-4}$~Myr age estimated by Zuckerman et al.  It is important to remember though that the uncertainties quoted on the age here are formal 
and are almost certainly a lower limit, with the true uncertainty larger than this quoted value.

\section{Discussion}

\subsection{Moving Group Member?}

The age and metallicity for PZ Tel~A measured in this work is slightly different from those previously published in the literature for this star.  The age we measure is greater than the age 
derived by other authors of 12$^{+8}_{-4}$~Myrs, based primarily on the notion that the PZ Tel system is part of the BPMG, only at the 1$\sigma$ level.  We find a mean age for PZ Tel~A of 
$\sim$24$\pm$3~Myrs, based on three different methods, two of which are in good agreement.  It is difficult to say whether this age can be transferred to the entire BPMG, if we assume 
that the PZ Tel system is a bonafide member of this kinematic group.  However, given the new Hipparcos reduction (\citealp{vanleeuwen07}) has altered the evolutionary status of PZ Tel~A and the 
star's high rotational velocity (which makes the measurement of accurate radial velocities very challenging), it may be necessary to reassess both if the PZ Tel system is an actual member 
of the BPMG and also the age of the BPMG itself.  On the other hand, this result is more in agreement with the age estimated for the BPMG by \citet{barrado99} of 20$\pm$10~Myrs, based on 
their discovery of three new M-dwarf members to this moving group, indicating that the BPMG is indeed older than 12Myrs.  

The metallicity ([Fe/H]) we measure for PZ Tel~A is found to be +0.05$\pm$0.20~dex, more than 0.30~dex higher than the earliest estimates for this star (\citealp{randich93}; 
\citealp{rocha-pinto98}), however more in agreement with the metallicity quoted later by \citet{rocha-pinto00} of 0.16~dex.  It is interesting to note that this later measurement by 
Rocha-Pinto et al. was made photometrically, whereas the most recent photometric metallicity estimate from \citet{holmberg07} provides a very metal-poor photospheric metallicity 
abundance of -0.50~dex.  This highlights the uncertainties inherent when using photospheric colours and magnitudes to estimate the metallicity of stellar atmospheres, particularly for 
young stars that show evidence for variability.  We argue that our metallicity should be more robust than past measurements, given the steps we explained above to ensure we selected the 
best Fe lines in our analysis.  Indeed, lower metallicity values were found when we randomly selected all lines, but this systematically affected the solar metallicity also by shifting it to 
more metal-poor values.  This may be the reason for the metal-poor nature found by Randich et al., along with their use of older model atmospheres and more incomplete spectral line lists.  

The solar/metal-rich nature of the PZ Tel system agrees well with the notion that young stars are formed in more metal-rich environments given enrichment of the interstellar medium by 
past supernovae explosions.  In addition, it also ties in with the finding of a debris disk around PZ Tel, which would indicate there was an abundance of metals in the remnant disk of material 
leftover by the formation of PZ Tel~A, even though the current disk is estimated to contain only 0.3~Lunar-masses of material (\citealp{rebull08}).  We searched the literature for spectroscopic 
metallicity values measured for other BPMG members, yet none were found, particularly in the works of \citet{zuckerman01}, \citet{feigelson06} and \citet{ortega09}.  Therefore, if PZ Tel~A is a 
BPMG member then this indicates that the BPMG is a cluster of around solar metallicity, if all members do indeed share similar metallicities.  Work is underway to explore this question 
(\citealp{ivanyuk11}).

\subsection{Evolutionary Model Testing}

\begin{figure}
\vspace{4.5cm}
\hspace{-4.0cm}
\includegraphics{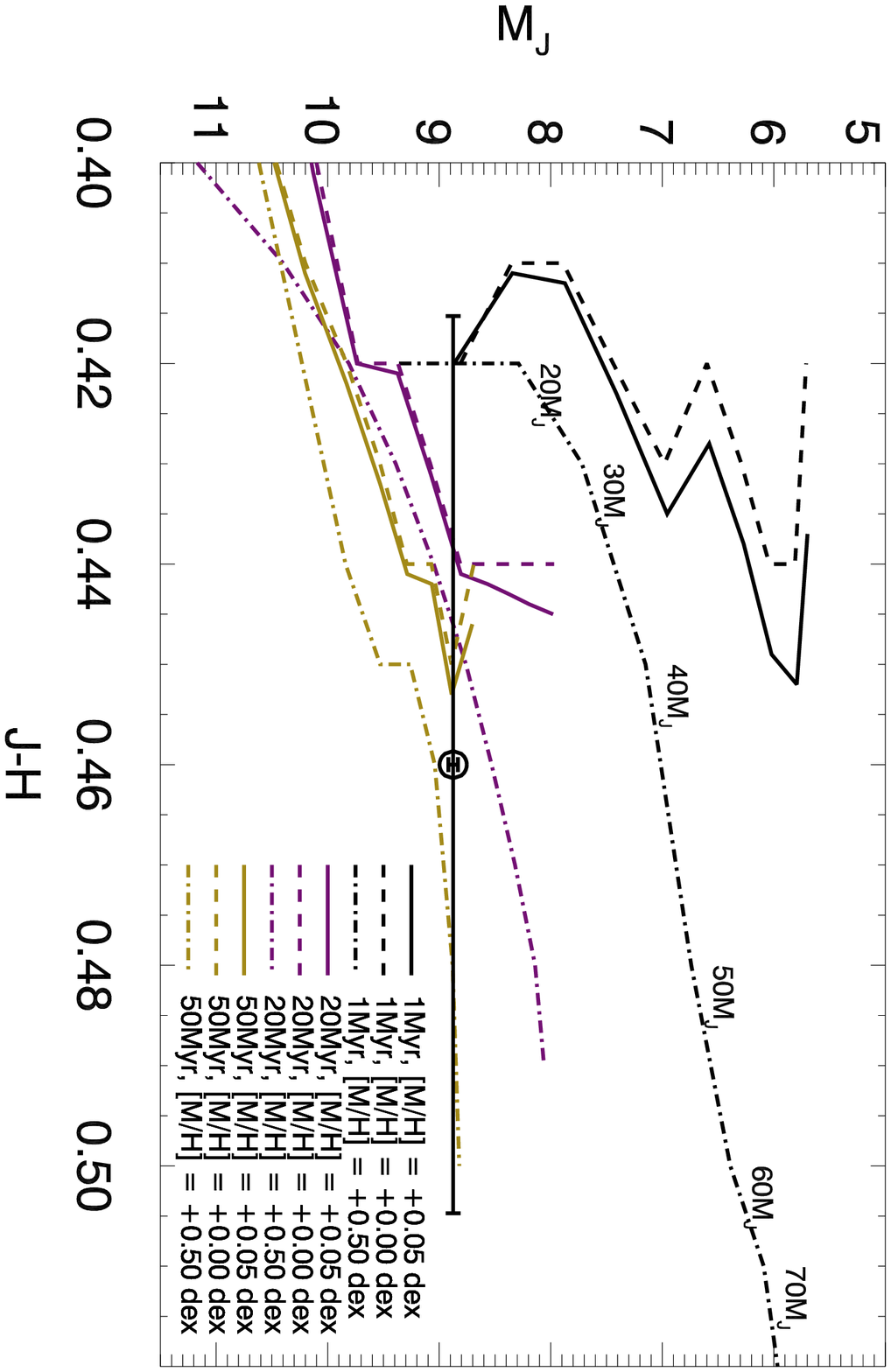}
\includegraphics{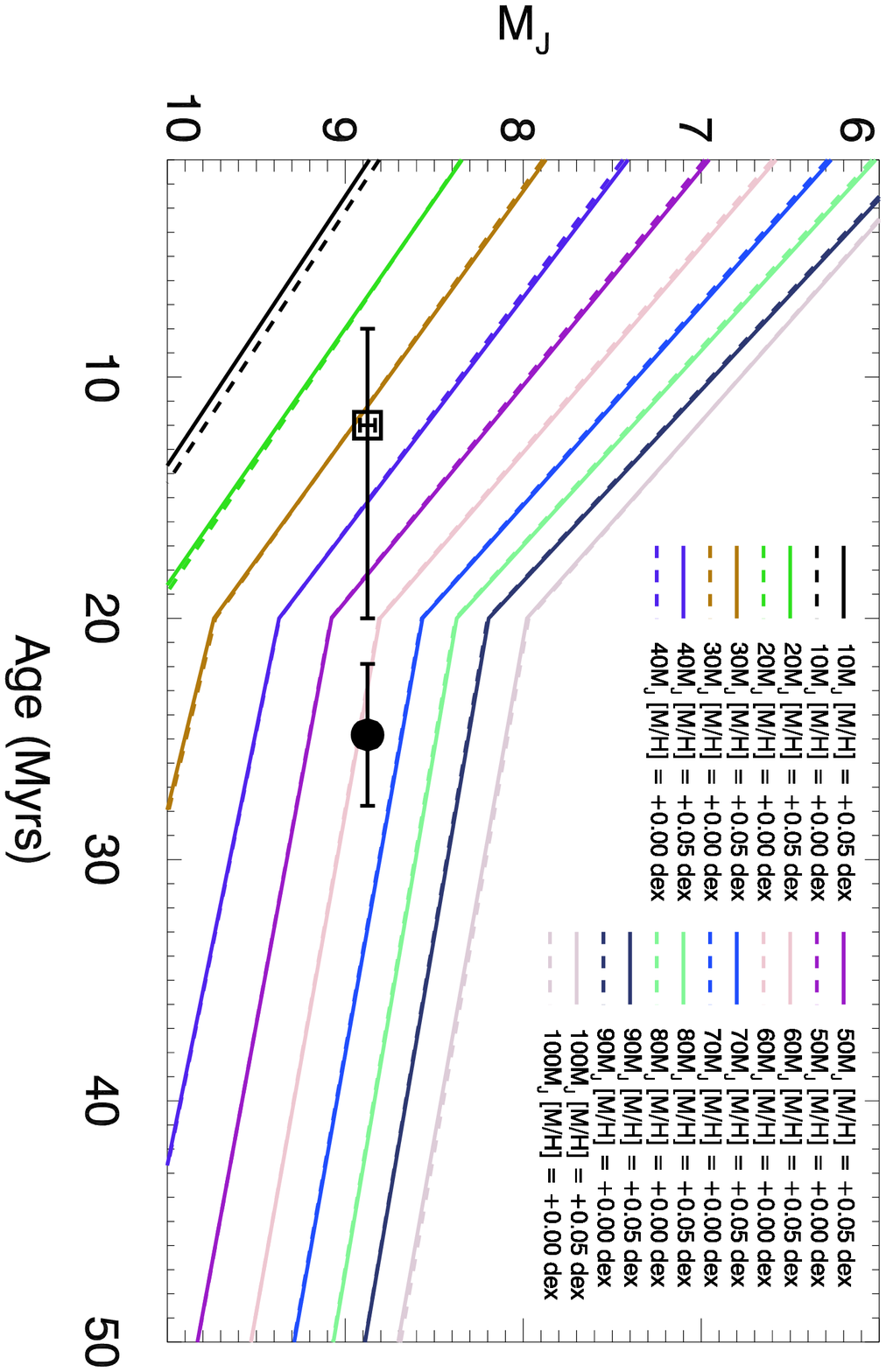}
\vspace{6.0cm}
\caption{The top panel shows a set of low-mass substellar evolutionary tracks on a colour-magnitude diagram.  The key highlights the ages as a function of colour and also we show 
solar metallicity models (dashed curves), our measured PZ Tel~A metallicity of +0.05~dex models (solid curves), and super-solar metallicity (+0.50~dex) models (dot-dashed curves).  
PZ Tel~B is represented by the open circle.  
The lower panel shows the change in absolute J-band magnitude as a function of age, for different substellar masses.  The solid and dashed curves represent the measured metallicity 
and solar metallicity models, respectively.  The filled circle is the position of PZ Tel~B, given our age estimate for the system, whereas the open square represents the age for the BPMG 
from \citet{zuckerman01}, assuming the PZ Tel system is a true member of the group.}
\label{evolutionary}
\end{figure}

Fig.~\ref{evolutionary} (upper panel) shows the position of PZ Tel~B on a colour-magnitude diagram.  The evolutionary calculations for such low mass objects are based on the Lyon 
stellar evolution code with input physics described in \citet{chabrier97} and \citet{baraffe98}. The outer boundary conditions between the interior and the non-grey atmosphere profiles 
are presented in detail in \citet{chabrier00b} as well as the treatment of dust in the atmosphere in \citet{allard01}.  We show them overplotted for metallicities of 0.00, +0.05 (metallicity 
of PZ Tel~A), and +0.50~dex, ages between 1-50~Myrs, and masses from 10-100M$_{\rm{J}}$.  The position of PZ Tel~B is represented by the open circle with its 
associated uncertainties, which in colour are fairly substantial.  The value is actually the mean values taken from \citet{biller10} and \citet{mugrauer10}.  

Firstly, we find that for young ages and 
such low-masses, increasing the metallicity from solar to +0.5~dex causes an increase in effective temperature at the level of $\sim$250K, at the upper mass limits for brown dwarfs 
($\sim$70M$_{\rm{J}}$).  Around PZ Tel~B we find this value is lower, at the level of $\sim$150K.  This highlights that increasing the metallicity of low-mass, young substellar 
objects has a significant effect on the bulk thermal properties of the system, particularly for a significant increase in metallicity at the +0.5~dex level.

Given the rather large uncertainties in the $J-H$ colour for PZ Tel~B, it does not facilitate a useful diagnostic at present to distinguish between different evolutionary tracks.  To 1$\sigma$ 
PZ Tel~B is in agreement with all the model tracks here.  Even so, currently the evolutionary tracks are not robustly tied to observations of substellar objects with well determined 
metallicities and gravities and therefore many more benchmark objects are needed to facilitate model constraints.

The lower panel in Fig.~\ref{evolutionary} shows the position of PZ Tel B as a function of our newly determined age, against the mean M$_{\rm{J}}$ magnitude from both \citet{biller10} 
and \citet{mugrauer10} (filled circle).  The evolutionary tracks are again plotted, with the solid curves representing the metal-rich model of +0.05~dex and the dashed curves representing 
the solar metallicity models.  We can firstly see that there is not a significant difference between the two evolutionary tracks, however for rigour we have used the +0.05~dex models in all 
our calculations.  We find PZ Tel~B to have a mass of 62$\pm$2~M$_{\rm{J}}$.  This value is around twice the quoted absolute mass values given in Biller et al. and Mugrauer 
et al., however we note again that the uncertainties on the age are formal and are certainly a lower limit and we quote a more realistic uncertainty from our Monte Carlo simulations that we 
explain below where we consider not just one fixed point in the parameter space but analyse our overall range of values.  The \teff and \logg measurements we obtain are 
2987$\pm$100~K and 4.78$\pm$0.10~dex, respectively.

In the figure we also show the position of PZ Tel~B given the Zuckerman et al. estimates for the age to the BPMG used in both Biller et al. and Mugrauer et al. to obtain their mass 
estimates for PZ Tel~B (open square).  The mass we derive from the models using this age estimate is 32$^{+24}_{-8}$M$_{\rm{J}}$, which runs from $\sim$24-56M$_{\rm{J}}$.  
Therefore, although the mass estimate we obtain given our new age measurements for the PZ Tel system is almost twice this value, the significance of the difference is only 
at the level of around 1.2$\sigma$.  The \teff and \logg measurements we obtain are 2377$^{+535}_{-340}$K and 4.68$^{+0.10}_{-0.07}$dex, respectively. 

 We note that \citet{simon11} have recently measured the diameters of two BPMG members HIP560 and HIP21547 using interferometric methods, and they find that evolutionary models would systematically overestimate the 
masses of these companions by around 0.2~M$_{\odot}$ and hence the ages are older by around 5~Myrs, both due to the effects of gravitational darkening.  They find an age of 13$\pm$2~Myrs 
for the BPMG.  If we take these systematic offsets and apply them to PZ Tel~A, we get a mass and age for this star of 1.03$\pm$0.04~M$_{\odot}$ and 19.83$\pm$2.94~Myrs.  This gives 
rise to a mass, \teff, and \logg of 57$^{+2}_{-10}$~M$_{\rm{J}}$, 2923$\pm$150~K, and 4.78$\pm$0.10~dex for PZ Tel~B, still placing it securely above the planetary limit.

\subsection{Monte Carlo Analysis}

\begin{figure}
\vspace{5.0cm}
\hspace{-4.0cm}
\includegraphics{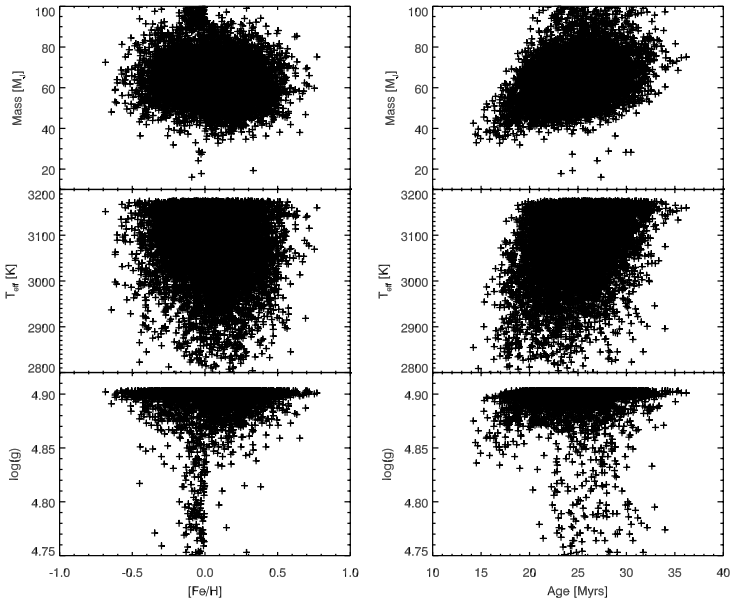}
\vspace{3.0cm}
\caption{The best fit parameters for mass (top), \teff (middle), and \logg (bottom) as a function of both metallicity (left column) and age (right column) for each 
of our 10000 random realisations of the possible values of metallicity, age, and $J$-band magnitude for PZ Tel~B.}
\label{mc_scatter}
\end{figure}

To better investigate the distribution of possible bulk properties for PZ Tel~B given our measured input parameters for PZ Tel~A, we perform a Monte Carlo 
(MC) analysis on the data.  To do this we randomly varied the input parameters age, metallicity, and $J$-band magnitude by their associated uncertainties, assuming a Gaussian 
model.  For each random realisation we then reinterpolate the evolutionary models and determine the best fit mass, \teff, and \logg. 

We performed 10000 random realisations in our simulation and the distributions for each are shown in Fig.~\ref{mc_scatter}.  In mass we see a well defined population as a 
function of metallicity, with an indication of a slight trend whereby metal-richness leads to less massive companions.  We also ran the same tests in the $H$ and $K_{s}$ bands to test 
if this trend towards lower-mass companions in metal-rich systems was found across the three near infrared bands, and not only the $J$-band.  We found the trend was apparent in all three 
bands, therefore current models suggest that in the near infrared, one is more sensitive to lower-mass companions 
around more metal-rich stars.  As a function of age we see the well understood trend between age and mass.  Taken together, higher 
metallicity and younger systems will produce lower-mass substellar objects, and hence future surveys could target the most metal-rich young stars to try to bias their searches 
towards lower-mass objects.  This scenario actually agrees with the prevalence of gas giant planets towards more metal-rich stars (\citealp{fischer05}), particularly when compared 
to field binary stars.

The distribution of both \teff and \logg show a number of features of this simulation, such as the truncation of the model boundaries, along with very non-symmetrical distributions.  
The \teff distributions generally follow the same trends as the mass distributions for both metallicity and age, since mass and \teff are heavily correlated at young ages for low-mass 
substellar brown dwarfs.  The distributions have parabolic boundaries, centered around the measured metallicity and age of the PZ Tel system, whereby the cooler models are spread 
more tightly around the measured values in comparison to the hotter models, which are generally found across a wide range in both metallicity and age.

As for \logg against metallicity and age, we find a dense population of values with a high surface gravity clustered around 4.9, and then a more collimated clustering towards 
lower gravities.  In the metallicity plane we see that for metal-poor values the \logg can have many solutions between 4.75 to 4.9, however the metal-rich models produce a 
clustering towards higher gravities.  A similar trend is seen in age, where young ages produce higher surface gravities, when compared to ages higher than the measured age of the 
PZ Tel system.  In fact, there are not many solutions in agreement with the measured \logg from the final position in age and metallicity.

\begin{figure}
\vspace{5.0cm}
\hspace{-4.0cm}
\includegraphics{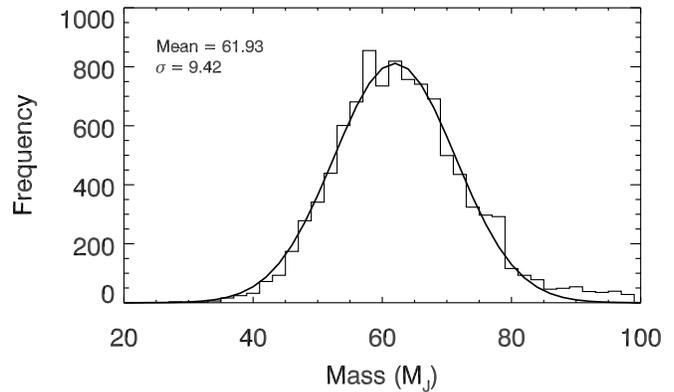}
\vspace{0cm}
\caption{Gaussian histogram distribution in mass for the Monte Carlo analysis performed to determine the bulk properties of PZ Tel~B.  Also overplotted is the best fit 
gaussian model to the data, along with the measured parameters in the top left.}
\label{mc_mass}
\end{figure}

The MC simulations were also used to better define our measured uncertainties for the parameters of PZ Tel~B.  An example of this is shown in Fig.~\ref{mc_mass} where we plot 
the frequency distribution of masses in histogram format for the entire set of 10000 random realisations.  The solid curve overplotted is the best fit gaussian model to the data 
and we show the measured parameters from this best fit model.  We find the mean of the mass distribution to be 62M$_{\rm{J}}$ and the standard deviation is 9M$_{\rm{J}}$.  Clearly this is 
much larger than the formal uncertainties of $\sim$2M$_{\rm{J}}$ quoted above for the measured solution and therefore we adopt this uncertainty for our mass measurements.  
We performed a similar analysis for the \teff and \logg and quote these uncertainties in Table~\ref{benchmark_table}.

\subsection{Benchmark Systems}

\begin{figure}
\vspace{5.0cm}
\hspace{-4.0cm}
\includegraphics{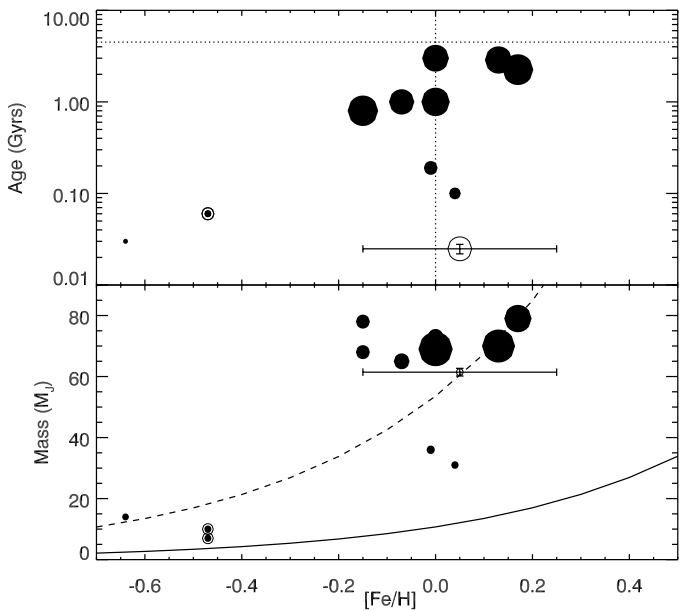}
\vspace{3.0cm}
\caption{The current distribution of age-metallicity benchmark binaries.  The top panel shows metallicity against age, with symbol sizes scaled by increasing mass 
of the companions.  The dotted cross-hairs represent the solar values in metallicity and age.  The lower panel shows metallicity against mass with the symbol sizes scaled by increasing age of 
the systems.  In both panels PZ Tel~B is shown as the open circle with its associated uncertainties, the benchmark systems are represented by the filled circles, and the ringed filled 
circles are the directly imaged planets around HR8799.  Also shown are formation models for core masses of 1\% (dashed curve) and 5\% (solid curve) of the total mass of the companions.}
\label{benchmarks}
\end{figure}

Fig.~\ref{benchmarks} shows the position of PZ Tel~B (open circles) with respect to other potential age-metallicity benchmark binary systems currently known (filled circles) that were taken 
from \citet{day-jones11b}.  In the top panel 
we show metallicity against system age in Gyrs and the lower panel shows metallicity against companion mass, in Jupiter-masses.  The characteristics of these benchmark systems are shown in 
Table~\ref{benchmark_table}, along with our calculated values for PZ Tel~B for direct comparison.  

The position of PZ Tel~B in metallicity and age space reveals just how this object extends the benchmarks into a new part of the parameter space.  PZ Tel~B is far from any other 
benchmark binary object currently known, potentially giving a new insight into 
the physics of solar metallicity, young brown dwarfs, once a spectrum of the object is obtained.  The symbol size representing each object is scaled to the mass of the most massive 
benchmark object, HD89744~B.  The mass of PZ Tel~B is around the median of the other benchmarks, which again means PZ Tel~B could be an interesting test case for 
studying low-mass atmospheric physics.  However, only $\eta$ Cancri~B has colours, or a spectral type, approaching that of PZ Tel~B, even though this object is at the opposite end 
of the age distribution.  We do note though that $\eta$ Cancri~B is likely an unresolved LT binary and therefore the properties for this object may not be accurate.

In this plot we also show the position of the Sun, in both metallicity and age, marked by the dotted lines, where the measured values are where these 
lines cross.  This allows us to see how the benchmarks compare to the Sun.  All benchmarks are younger than the Sun, which represents a strong bias towards younger objects, given 
they are much more luminous for a given mass, with AB~Pic~B and PZ Tel~B much younger than the Sun and the other benchmark binaries.  Also we see that most are fairly close to the 
solar metallicity by around $\pm$0.2~dex, including PZ Tel~B, except the extremely metal-poor brown dwarf binary, AB~Pic~B.  Taken all together, it seems that we have 
covered a large fraction of the benchmark metallicity-age space, except for, 1) all metallicities with ages older than the Sun ($\ge$4~Gyrs), 2) metallicities less than -0.2 with 
ages above $\sim$0.1~Gyrs, and 3) young ages ($\le$1~Gyrs) with solar/super-solar metallicity ($\ge$0.0~dex), which is where PZ Tel~B finds itself positioned.

The lower panel shows the mass range for the population against metallicity and there is a clear clustering of objects with high brown dwarf masses and in more solar/metal-rich 
environments.  Higher masses would be favoured as there is a strong bias towards such objects, given they appear brighter on the sky for a given distance than lower mass objects 
and this can go some way to explaining this mass clustering.  In this case the symbol sizes are scaled by age, from the oldest benchmark $\eta$ Cancri~B.  Interestingly, PZ Tel~B is found 
clustered around a number of higher mass brown dwarfs, both in metallicity and mass space, however it is clearly much younger than the others.  PZ Tel~B actually completes a missing 
age piece of the evolutionary scale in this mass and metallicity parameter space, meaning this region is reasonably well sampled in comparison to other regions.  Probably the most 
important point here is that there is a lack of low-mass objects across all metallicities, and the three objects with masses less than 60M$_{\rm{J}}$, are found to be very young, 
which if PZ Tel~B is really younger than our measurement of $\sim$24~Myrs, then it will drop into this regime, but again would be a very young object.

We also note that for these binaries there may also be a formation effect at play here too when looking at mass, depending if the companion has formed through direct gravitational 
collapse of the remnant disk, or if it has formed through a core accretion process, more akin to planet formation.  Core accretion models generally do have difficulty forming massive 
objects, however on rare occasions, models do predict such large bodies can form through the planet formation process.

We overplot two core accretion based models, using the relationship discussed in $\S$~1.1.  We have made some broad assumptions here, for instance, the companions all have a fixed envelope 
to core mass fraction.  The two curves shown represent 5\% (solid) and 1\% (dashed) core mass fractions, and we have extended the 1\% model up to the brown dwarf boundary.  Physically this 
may be unrealistic as the core accretion models have problems building such large objects (see \citealp{mordasini09a}).  Fragmentation of the remnant disk might be a better way to form these 
higher mass companions (see \citealp{stamatellos09}).  However, these basic models do show that when there is a high fraction of metals in a proto-planetary disk, and extremely large cores 
can be formed, if we input realistic core-to-envelope ratios of only a few percent we can go a long way to explaining the distribution of benchmark binaries in metallicity-mass space.

In both the upper and lower plots the filled circle encased by the rings mark the position of the directly imaged planets HR8799~$b,c$~and~$d$ (\citealp{marois08}).  If these wide orbiting 
planets were formed by core accretion, along with the benchmark brown dwarf binaries, then it is worth comparing the properties of these planets with the brown dwarfs.  Of course, again we 
have the biases, particularly in age, where these planets were only able to be imaged with current technology since they are so young, and hence bright.  However, in the metallicity-mass 
parameter space, these planets reside between the two formation models, along with a few of the benchmark binaries, including PZ~Tel~B. 

\subsection{Benchmark Colours}

\subsubsection{J-H vs H-K}

\begin{figure}
\vspace{5.0cm}
\hspace{-4.0cm}
\includegraphics{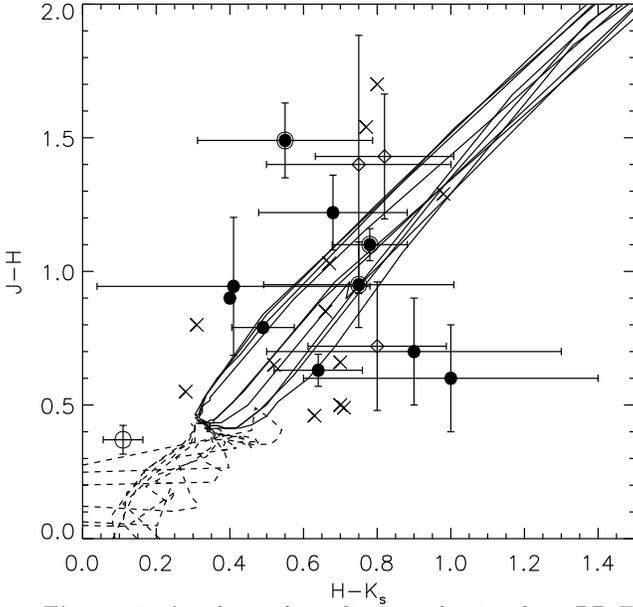}
\vspace{2.6cm}
\caption{A colour-colour diagram showing how PZ Tel~B (open circle) compares against other benchmark binaries currently known (filled circles).  Also shown are the planets 
around HR8799 (open diamonds) and a list of free-floating planetary mass objects (crosses).  The filled circles with rings around them show objects with masses and ages close to 
PZ Tel~B.  The solid and dashed curves are DUSTY and COND evolutionary models respectively.}
\label{col_col}
\end{figure}

Late-M stars are known to be affected by emergent dust in their atmospheres at \teff's below the temperature of PZ Tel~B (\citealp{jones97}).  They also exhibit features in their near 
infrared spectra such as water bands around 1.4, 1.8 and beyond 2.4$\mu$m, CO bands at around 2.35$\mu$m, FeH bands at 0.99 and 1.2$\mu$m, strong $J$-band potassium absorption.  
We show the position of PZ Tel~B in comparison to our other benchmark systems on a colour-colour diagram ($H-K_{s}$ against $J-H$) in Fig.~\ref{col_col}.  Also shown for comparison 
are the DUSTY (\citealp{chabrier00}) and COND (\citealp{baraffe03}) evolutionary tracks, represented by the solid and dashed curves, respectively.

First of all, we find that PZ Tel~B is much 
bluer in both colour bands than the other benchmark binaries we have discussed, including the binaries with masses and ages near that of PZ Tel~B.  It is also bluer than some 
young planetary-mass M dwarfs from the Orion Cluster that we also highlight (\citealp{weights09}), at least bluer in the $H-K_{s}$ colour index.  This could be an affect related to 
the difference in metallicity between PZ Tel~B and the Orion Cluster objects since some estimates for the metallicity of the cluster place it below the solar value (see \citealp{odell01}), 
however later estimates claim a more solar metallicity value (\citealp{dorazi09}).

These low-mass and young free-floating planetary-mass objects are useful to interpret PZ Tel~B, since they are of similar age and \teff, and they help us to realise that 
PZ Tel~B could be an M, L, or T-dwarf or even a planet.  Again this raises questions on formation scenarios for young and low-mass objects, but questions over the true physical 
nature of PZ Tel~B will require spectroscopic follow-up.

\citet{leggett01} note that when there is no dust present in the atmospheres of M dwarfs, the water bands are expected to become deeper and exhibit increasingly steep wings with 
decreasing \teff.  However, in the presence of dust, the atmosphere is heated and the bands become wider and shallower.  We have shown that PZ Tel~B is not a metal-poor companion through 
association to the host star PZ Tel~A, and hence PZ Tel~B can be expected to host a dusty atmosphere.  In particular, we predict the water bands for this companion to be 
shallow and broad in comparison to more metal-poor field M dwarfs of a similar \teff. 

Another opacity source affecting such cool atmospheres is that of collisionally induced hydrogen absorption (\citealp{linsky69}).  Suppressed $K$-band flux has been 
attributed to this and has been used in colour selection criteria to select unusual substellar objects (e.g. \citealp{murray11}).  However, since PZ Tel~B does not appear to be significantly redder 
in the $H-K_{s}$ colour band compared with the other young M dwarfs we show, it appears that this type of absorption is not a significant source of opacity in the atmospheres of young 
and low-mass brown dwarfs like PZ~Tel~B.  Given that we only have one data point here at present, this may potentially provide a future avenue of research.

Although PZ~Tel~B is bluer in both colours in comparison with the companion planets of HR8799 (open diamonds), the free-floating planetary-mass objects exhibit similar 
colours to these planets.  This shows that the evolutionary properties of young low-mass brown dwarfs and high-mass planets are similar, validating their use as benchmarks to 
better understand the physics of gas giant planets.  

\subsubsection{Metallicity vs Near-IR Photometry}

\begin{figure}
\vspace{5.0cm}
\hspace{-4.0cm}
\includegraphics{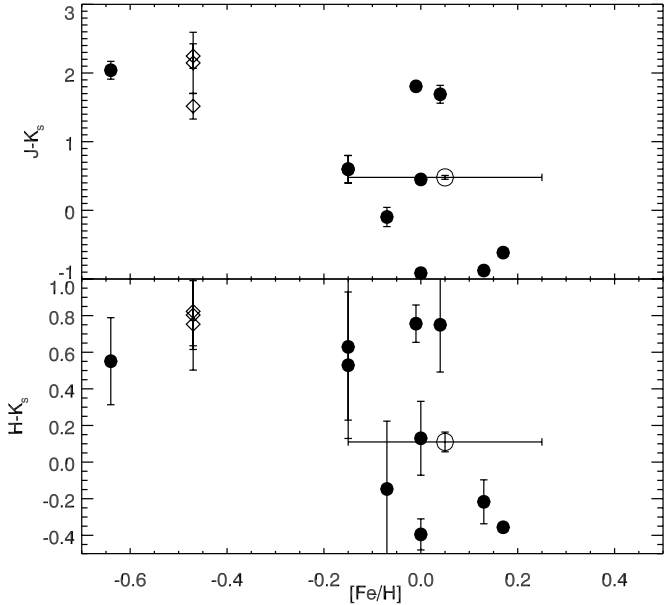}
\vspace{3.0cm}
\caption{The distribution of near infrared colours $J-K_{s}$ (top) and $H-K_{s}$ (bottom) as a function of increasing metallicity for PZ Tel~B (open circle), the comparison benchmark 
binaries (filled circles), and the planets orbiting HR8799 (open diamonds).  All object colours were scaled using evolutionary models to the age and mass of PZ Tel~B for comparison.  
Uncertainties are included where they were published in the literature.}
\label{met_col}
\end{figure}

Since metallicity can have a strong impact on the atmospheric opacities of cool substellar objects, studying how colours evolve as a function of metallicity can allow a 
deeper understanding of the physics and interactions ongoing within these objects.  In Fig.~\ref{met_col} we show the distribution of $J-K_{s}$ and $H-K_{s}$ colours 
against changing metallicity.  Given that substellar colours evolve as a function of time and as a function of mass, each object's colour was scaled to the age and mass of 
PZ Tel~B using the DUSTY evolutionary models (\citealp{chabrier00}).  This scaling allows us to attempt to isolate the effects of metallicity on the atmospheric properties 
of these companions.

At first glance there maybe a trend between the metallicity and the broadband colours of these benchmark companions, even when including the planets around HR8799
(open diamonds). A possible anti-correlation is present, whereby an increase in metallicity tends to decrease the colour index, at least in the metal-rich regime. 
The Pearson rank correlation coefficient for the $J − K_{s}$ data is -0.73, suggestive of a strong trend between these two parameters. Such a trend would indicate a 
turnover is present, since the metal-poor L sub-dwarfs also exhibit bluer $J − K_{s}$ colours (e.g. \citealp{burgasser09}; \citealp{lodieu10}).  If this is the case, then 
it would provide an explanation for at least some of the peculiar blue L-dwarfs discussed in length in \citet{kirkpatrick10}.  Objects like 2M1711+4028~B, which have been 
shown to have solar metallicity (\citealp{radigan08}) and blue colours, would naturally be explained if there is a negative trend between near infrared colours and metallicity 
in the more metal-rich regime.

If at least part of the correlation discussed is real, it could be more important in the bluer bands than in the redder bands, as it affects the $J − K_{s}$ colour more than it
affects the $H − K_{s}$ colour. The Pearson rank coefficient for the $H − K_{s}$ colour is -0.65, which still indicates a trend, but at a reduced level of significance.

\begin{figure}
\vspace{5.0cm}
\hspace{-4.0cm}
\includegraphics{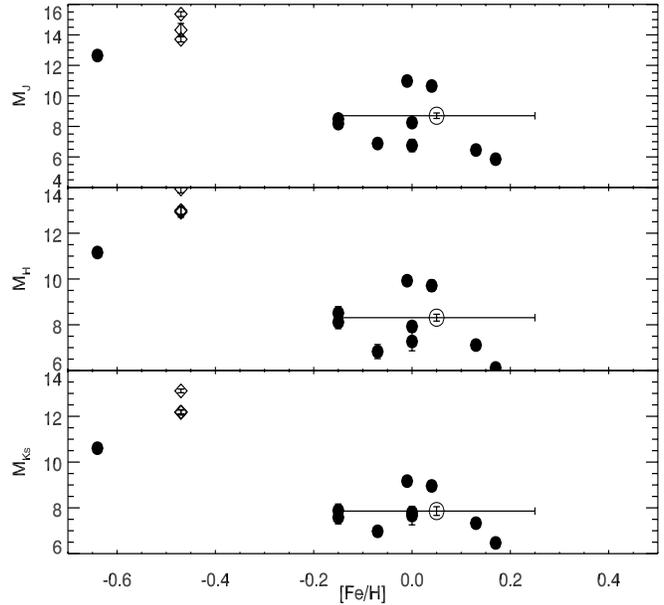}
\vspace{3.0cm}
\caption{The near infrared absolute magnitudes against metallicities for PZ Tel~B (open circle), the benchmark binaries (filled circles), and the planets around HR8799 (open diamonds).  
From top to bottom we show absolute $J$, $H$, and $K_{s}$ magnitudes, respectively.  The absolute magnitudes of the benchmarks have been scaled to the age and mass of PZ Tel~B 
for direct comparison.}
\label{met_mag}
\end{figure}

In Fig. 10 we show the absolute $J$, $H$, and $K_{s}$ magnitudes against metallicity to test if any gradients are apparent in the absolute fluxes that would support the possibility 
of a decreased level of dust or absorption is present.  Indeed, we do see a possible anti-correlation between the near infrared photometry and increasing metallicity, with the trend
appearing a little stronger in the $J$-band than the redder bands.

However, given that the two metal-poor data points are both young and of low-mass there could still be an age and mass correlation present which has not been fully corrected for due to 
model uncertainties. The models have known difficulties in reproducing the bulk properties of young and low-mass brown dwarfs. To test this further it is necessary to populate the
metal-poor regime with more older and more massive brown dwarf companions to better constrain this region of the parameter space.

\begin{landscape}
\begin{table*}
\vspace{5.0cm}
\tiny
\center
\caption{Potential metallicity benchmark binaries}
\label{benchmark_table}
\scalebox{0.85}{
\begin{tabular}{cccccccccccccc}
 &&&&&&&&&&&&& \\
\hline

UCD        &      SpT    &      Primary  &   D     & Sep        &   Age    &   Mass        &  Teff  &  \logg   &     [Fe/H] &    $J$     &  $J-H$  &  $J-K_{s}$ &  Ref\\
ID         &             &               &   pc    & AU         &   Gyr    &   M$_{Jup}$    &   K    &  dex      &      dex  &    2MASS   &  2MASS  &   2MASS & \\

\hline

Gl 417~B             &        L4.5            & G0V             & 21.93$\pm$0.21$^{a}$             &     $\sim$2000     &      0.08-0.3  &    36$\pm$15     &    1600-1800      &        -  &  -0.01$^{e}$       &     14.57$\pm$0.04 &     1.10$\pm$0.06 & 1.88$\pm$0.06 &  1, 2\\
GJ1048~B             &        L1              & K2Vk             & 21.27$\pm$0.43$^{a}$   &    250     &   $<$1.0   &   55-75     &   1900-2200    &  -  &   -0.07     &  13.669$\pm$0.160 &   0.944$\pm$0.258    & 1.354$\pm$0.140   & 3\\
LHS102~B             &        L8              & M4V             & 9.56$\pm$0.49$^{a}$             &     180            &      $>$1.0    &    73$\pm$10     &    1900-2000      &   6.0$^{d}$ &  0.00 $^{d}$      &        13.30$\pm$0.09   &  1.22$\pm$0.14$\dagger$  & 1.90$\pm$0.07 & 4,5\\
2M1711+4028~B         &        L4$^{+2}_{-1.5}$ & M4.5V          &  21.2$\pm$3.9$^{a}$     &     135$\pm$25     &     1-5        &    69$^{+8}_{-15}$  & 1700$^{+210}_{-250}$ &        -     &     -1.0$^{e}$    &  15.00$\pm$0.06  &  0.79$^{c}$ &   1.28$^{c}$ & 6\\
HD89744~B            &        L0              & F7IV-V          &  39.43$\pm$0.48$^{a}$               &     2460           &     1.5-3.0    &    77-80            & 2000-2200      &        -                    & 0.17$^{e}$        &      14.90     &     0.90      &    1.30 & 7, 8\\
AB Pic~B             &        L0-L1           & K2V             & 47.3$^{+1.8}_{-1.7}$     &      250-270       &    0.03        &      13-14    &    2000$^{+100}_{-300}$ &    4.0$\pm$0.5   &          -0.64$^{e}$       & 16.18$\pm$0.10  & 1.49$\pm$0.14   & 2.04$\pm$0.13 & 9\\
HD 130948~B          &        L0-L4           & G2V             &  18.17$\pm$0.11$^{a}$             &     $\sim$48       &      $<$0.8    &    $<$78      &       1950$\pm$250    &       -           &         -0.15$^{e}$        &      13.9$\pm$0.2$^{b}$  &    0.7$\pm$0.2$^{b}$  & 1.6$\pm$0.2$^{b}$ & 5, 10\\
HD 130948~C          &        L0-L4           & G2V             &  18.17$\pm$0.11$^{a}$             &     $\sim$48       &      $<$0.8    &    $<$68      &       1950$\pm$250    &       -           &         -0.15$^{e}$        &      14.2$\pm$0.2$^{b}$  &    0.6$\pm$0.2$^{b}$  & 1.6$\pm$0.2$^{b}$ & 5, 10\\
$\eta$ Cancri~B         &        L0.5$\pm$1.5    & K3III           &  91.49$\pm$3.35$^{a}$             &     14497-15581    &     2.24-3.50  &    68-72      &       1920$\pm$100    &  5.30-5.50        &     0.13$\pm$0.06      &        17.78$\pm$0.06$\dagger$ & 0.63$\pm$0.06$\dagger$  & 1.27$\pm$0.06$\dagger$ &  11\\ 
CD-352722B	&	   L4$\pm$1	&    M1Ve     &    21.3$\pm$1.4    &    67$\pm$4    &  0.1$\pm$0.05	&   31$\pm$8    &  	    1800$\pm$100  &   4.5$\pm$0.5  &  0.04$\pm$0.50  & 13.69$\pm$0.11$\dagger$	 & 0.95$\pm$0.16 & 1.70$\pm$0.13$\dagger$  & 12\\
PZ Tel~B            &        M6              & G9.5            &  51.49$\pm$2.60$^{a}$         &     $\sim$16       &   0.024$\pm$0.003 & 62$\pm$9 & 2987$\pm$100 &   4.78$\pm$0.10 &    0.05$\pm$0.20        &     12.26$\pm$0.14  &    0.39$\pm$0.17  &  0.84$\pm$0.21 & 13, 14 \\

\hline

\multicolumn{13}{l}{$^{a}$: parallax of primary from \citep{vanleeuwen07} or \citep{perryman97}, $^{b}$: Combined value from unresolved properties of a UCD+ UCD binary system.}\\
\multicolumn{13}{l}{$^{c}$: synthetic colours calculated from the spectrum [6],$^{d}$: [4], $^{e}$: \citet{day-jones11b}}\\
\multicolumn{13}{l}{1: \citet{kirkpatrick01}, 2: \citet{lachaume99}, 3: \citet{gizis01}, 4: \citet{leggett02}, 5: \citet{nordstrom04}, }\\ 
\multicolumn{13}{l}{6: \citet{radigan08}, 7: \citet{burgasser05}, 8: \citet{wilson01}, 9: \citet{chauvin05}, } \\ 
\multicolumn{13}{l}{10: \citet{potter02}, 11: \citet{zhang10}, 12: \citet{wahhaj11}, 13: \citet{biller10}, } \\
\multicolumn{13}{l}{14: \citet{mugrauer10}. $\dagger$MKO-2MASS conversion from \citet{hodgkin09}} \\

\hline 
\end{tabular}
}
\medskip
\end{table*}
\end{landscape}

\section{Summary}

From our age and abundance analysis of the pre-main sequence star PZ Tel~A, we have been able to place the low mass companion to this star (PZ Tel~B) as the youngest 
benchmark object with both an independently measured age and metallicity.  We find an age for the PZ Tel system older than previous measurements and our new method 
to determine the metallicity of rapidly 
rotating stars has shown that this system is also more metal-rich than previously thought.  Given the young age of the brown dwarf being around 24$\pm$3~Myrs, PZ Tel~B can 
provide an anchor point to scale evolutionary models at young ages, particularly when used in conjunction with older benchmarks like Wolf 940~B (\citealp{burningham09}).  

From our age estimates of the host star, we have measured bulk properties for PZ Tel~B by interpolation within a grid of evolutionary models.  We find PZ Tel~B to be more massive and 
hotter than previous estimates.  The mass we find ($\sim$62$\pm$9M$_{\rm{J}}$) places it above the current best estimates for the dividing line between planetary-mass companions 
and brown dwarf binary companions.  However, we show using a basic core accretion derived model that when remnant proto-planetary disks contain a high metal fraction, 
and can form large cores (around a few x10$^{2}$M$_{\oplus}$), core-to-envelope ratios of only a few percent can describe well most of the current benchmark binaries in 
metallicity-mass space.  

We ran Monte Carlo simulations to explore the parameter space of our evolutionary models in mass, \teff, and \logg as a function of metallicity and age.  In general we found that 
more metal-rich systems tend towards lower-mass companions.  Therefore, current models suggest that one could increase their sensitivity to lower-mass companions in the near 
infrared by searching around more metal-rich primaries.

We also tested if metallicity has an influence on the near infrared colours of brown dwarfs through varying dust opacity and found a possible trend that suggests that metal-rich 
atmospheres exhibit increased near infrared flux, particularly in the bluer near infrared bands.  We accounted for age and mass effects using model corrections and identified a possible 
metallicity trend.  However, due to mass/age biases in our benchmark sample this apparent metallicity trend may be caused by uncertainties in our model corrections, and an 
improved sample is needed.  Finally, this paper can serve as a blueprint for future analyses of long period exoplanetary systems that host gas giants like Jupiter 
(e.g. \citealp{jones10}) that will be directly imaged by instruments like the GPI (\citealp{macintosh06}) and SPHERE (\citealp{beuzit06}).

\section*{Acknowledgments}

We acknowledge the comments from the anonymous referee that have helped to make this a better manuscript.  We also thank Phil Lucas for some useful suggestions.  
JSJ acknowledges funding by Fondecyt through grant 3110004 and partial support from Centro de Astrof\'\i sica FONDAP 15010003, the  GEMINI-CONICYT FUND 
and from the joint committee ESO-Government of Chile grant.  ADJ is funded by a Fondecyt postdoctorado, under project number 3100098.  DJP, HRAJ, YP, OI and LY are supported by RoPACS, 
a Marie Curie Initial Training Network funded by the European Commission's Seventh Framework Program.  MTR acknowledges partial support from CATA PB-06 and 
FONDAP 15010003.  Based on observations made with the European Southern Observatory telescopes obtained from the ESO/ST-ECF Science Archive Facility.

\bibliographystyle{mn2e}
\bibliography{refs}

\end{document}